\documentclass[times, 10pt]{elsarticle}

\usepackage{amssymb}
\usepackage{amsmath}

\usepackage{url}
\usepackage{color}
\usepackage{multicol}
\usepackage{multirow}
\usepackage{bbding}
\usepackage{pifont}
\newcommand{\revisioncolor}{black}
\usepackage{hyperref}
\hypersetup{
    hypertex=true,
    colorlinks=true,
    linkcolor=green,
    anchorcolor=blue,
    citecolor=green
}

\journal{Pattern Recognition}

\begin{document}

\begin{frontmatter}




\title{TransVFC: A Transformable Video Feature Compression Framework for Machines}

\affiliation[label1]{organization={Institute of Information Science, Beijing Jiaotong University},
            city={Beijing},
            postcode={100044},
            country={China}}

\affiliation[label2]{organization={Visual Intelligence +X International Cooperation Joint Laboratory of MOE},
            city={Beijing},
            postcode={100044},
            country={China}}

\affiliation[label3]{organization={School of Computer and Communication Engineering, University of Science and Technology Beijing}, 
            city={Beijing},
            postcode={100083}, 
            country={China}}
            
\affiliation[label4]{organization={School of Computer Science and Engineering, Nanyang Technological University}, 
            postcode={639798}, 
            country={Singapore}}

\author[label1,label2]{Yuxiao Sun}
\ead{22110081@bjtu.edu.cn}
\author[label1,label2]{Yao Zhao}
\ead{yzhao@bjtu.edu.cn}
\author[label1,label2]{\corref{cor1}Meiqin Liu}
\ead{mqliu@bjtu.edu.cn}
\author[label3]{Chao Yao}
\ead{yaochao@ustb.edu.cn}
\author[label1,label2]{Huihui Bai}
\ead{hhbai@bjtu.edu.cn}
\author[label1,label2]{Chunyu Lin}
\ead{cylin@bjtu.edu.cn}
\author[label4]{Weisi Lin}
\ead{wslin@ntu.edu.sg}

\cortext[cor1]{Corresponding author}

\begin{abstract}

\footnotetext{\color{blue}{This is the author’s version (AAM version) of the work accepted for publication in Elsevier Pattern Recognition.
The final version of record is available at: \url{https://doi.org/10.1016/j.patcog.2025.112091}}}

Currently, an increasing number of video transmissions are focusing primarily on downstream machine vision tasks rather than on human vision. While the widely deployed human visual system (HVS)-oriented video coding standards such as H.265/HEVC and H.264/AVC are efficient, they are not the optimal approaches for video coding for machines (VCM) scenarios, leading to unnecessary bitrate expenditures.  
Academic and technical explorations within the VCM domain have led to the development of several strategies; however, conspicuous limitations remain in their adaptability to multitask scenarios. 
To address this challenge, we propose a Transformable Video Feature Compression (TransVFC) framework. It offers a compress-then-transfer solution and includes a video feature codec and feature space transform (FST) modules. 
In particular, the temporal redundancy of video features is squeezed by the codec through a scheme-based inter-prediction module. Then, the codec implements perception-guided conditional coding to minimize spatial redundancy and help the reconstructed features align with the downstream machine perception process.
Subsequently, the reconstructed features are transferred to new feature spaces for diverse downstream tasks by the FST modules. To accommodate a new downstream task, only one lightweight FST module needs to be trained, avoiding the need to retrain and redeploy the upstream codec and downstream task networks. Experiments show that TransVFC achieves high rate-task performance for diverse tasks at different granularities. 
We expect our work to provide valuable insights for video feature compression in multitask scenarios. The codes are available at \url{https://github.com/Ws-Syx/TransVFC}.

\end{abstract}



\begin{keyword}
Neural video compression \sep Feature compression \sep Video coding for machines, Intermediate feature


\end{keyword}

\end{frontmatter}



\section{Introduction}

Digital videos play a crucial role in our lives, constituting a significant portion of the information consumed daily. For videos aimed at the human visual system (HVS), such as movies and short clips, preserving visual details perceptible to humans during the compression process is essential. Moreover, videos collected for machine vision tasks, such as surveillance~\cite{PANDEESWARI2025111349} and facial recognition~\cite{facial1}, do not require all of their visual details to be preserved~\cite{scalable_choi, scalable_dcc}. Recently, neural video compression frameworks for the HVS have evolved significantly and now offer excellent video compression performance~\cite{dcvc_fm, dcvc_dc, dcvc_hem}. However, a comprehensive exploration of neural-based video coding for machines (VCM) remains nascent. 

\begin{figure}[!t]
\centering
\includegraphics[width=0.8\linewidth]{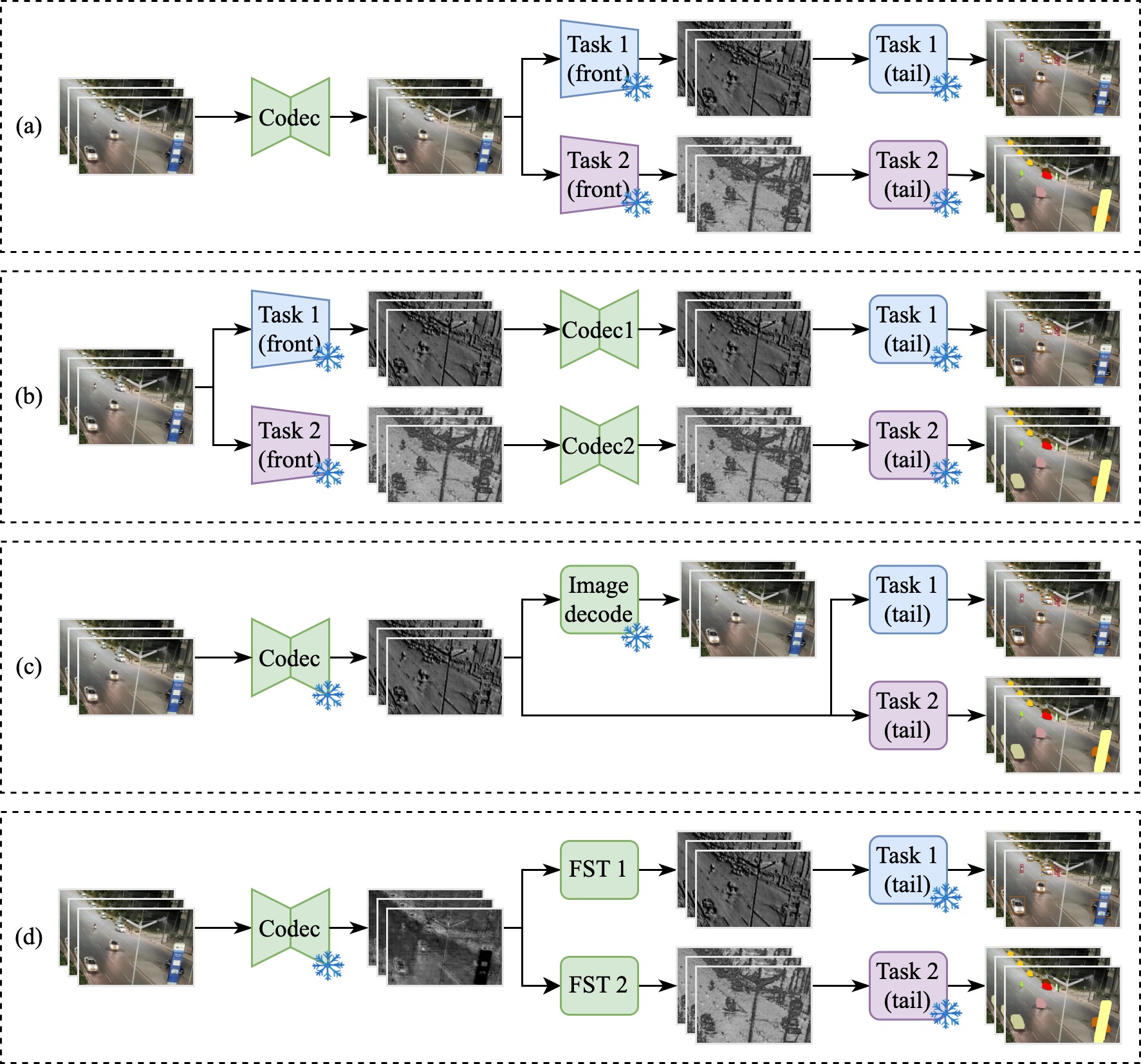}
\caption{
A comparison among different pipelines in VCM scenarios. ``Task(front)'' represents the shallow layers of the downstream task network, ``Task(tail)'' denotes the rest of the downstream task network, and the snowflake symbol represents ``module weights are frozen''.
(a) Videos are compressed by a hybrid or neural-based codec~\cite{h264, h265, smc, smc++} and analyzed by downstream networks.
(b) Intermediate features are extracted by the shallow layers of the task network and compressed by a specific-optimized video feature codec~\cite{misra2022video, shao2020bottlenet++}.
(c) The intermediate features for frame reconstruction are used to perform machine vision tasks, and the whole downstream task network is optimized~\cite{sheng2023lvvc}.
(d) Our framework uses a video feature codec for continuous feature transmission, then transfers the reconstructed features to various downstream tasks via lightweight feature space transform (FST) modules.
}
\label{fig_1}
\end{figure}

HVS-oriented video codecs, such as H.265/HEVC~\cite{h265} and H.264/AVC~\cite{h264}, are frequently employed to compress videos for downstream analysis, as shown in Figure~\ref{fig_1}(a). However, these approaches encounter two limitations in VCM scenarios. 
First, these compression frameworks focus on minimizing pixel-domain and HVS-related distortion, such as the peak signal-to-noise ratio (PSNR) and the multi-scale structural similarity index measure (MS-SSIM), rather than meeting the specific needs of machine vision applications, which is a suboptimal approach for machine vision. 
Second, machine vision tasks usually require only a subset of image content~\cite{scalable_choi, cta_preprocess3}. For example, indiscriminately transmitting the background of an image for the downstream image classification task leads to bitrate waste. More tailored approaches are needed in machine-centric scenarios.

Some studies have delved into the analyze-then-compress (ATC) paradigm to solve the above problems. The paradigm begins by extracting features from images, followed by feature compression for specific downstream tasks~\cite{yamazaki2022deep}. 
To enhance the versatility, some studies~\cite{scalable_choi} focus on mining the generalization of intermediate features across various downstream tasks. Nonetheless, the above advancements only cater to intra-compression, and do not address the temporal redundancy in continuous features. 
For video feature compression, one strategy~\cite{misra2022video, shao2020bottlenet++} entails optimizing a video feature codec by the specific downstream loss, as shown in Figure~\ref{fig_1}(b). Alternatively, another strategy~\cite{sheng2023lvvc} focuses on freezing the codec while fine-tuning the entire downstream task network, as described in Figure~\ref{fig_1}(c).
However, these approaches require retraining and redeploying either the upstream codec or the downstream machine vision networks to accommodate new downstream tasks, thus costing more computational resources and limiting their scalability in real-world applications.

To achieve better scalability and versatility in multitask scenarios, we propose a Transformable Video Feature Compression (TransVFC) framework that offers a compress-then-transfer solution. As illustrated in Figure~\ref{fig_1}(d), our proposed framework contains an innovative neural-based video feature codec and diverse lightweight feature space transform (FST) modules. Specifically, the codec employs a scheme-based inter-prediction module to squeeze the temporal redundancy of video features and form a coarse compensated feature. Furthermore, it conducts perception-guided conditional coding for fine reconstruction and helps the reconstructed feature align with the downstream machine perception process. Subsequently, the reconstructed features are transferred to other feature spaces of diverse downstream machine vision tasks via the FST modules. For any new downstream task, only one lightweight FST module must be trained instead of retraining and redeploying the upstream codec or the networks of the downstream tasks. Experiments are conducted on three machine vision tasks at different granularities. The results demonstrate that the proposed TransVFC outperforms the state-of-the-art (SOTA) neural codecs on all downstream tasks and outperforms VTM-23.1~\cite{vtm} on video instance segmentation and object detection. The contributions of this study are as follows. 

\begin{itemize}

    \item We propose a novel Transformable Video Feature Compression (TransVFC) framework. It comprises two components: a video feature codec and diverse feature space transform modules, offering a scalable and deployable VCM solution. 
    
    \item We introduce an innovative neural-based video feature codec to squeeze the redundancy encountered in the feature domain. It includes a scheme-based inter-prediction module and a perception-guided conditional coding module.
    
    \item We design a lightweight feature space transform module that transfers intermediate features to diverse downstream tasks in a highly scalable way. The experimental results validate the scalability and effectiveness of TransVFC across multiple downstream machine vision tasks of varying granularities.

\end{itemize}

\section{Related Works}

\subsection{Neural Video Compression}

Most of the existing neural video compression methods follow the motion-then-residual paradigm~\cite{dcvc_dc, dcvc_fm, ibvc} and mainly include inter-prediction and residual (i.e. context) compression. Lu \emph{et al.}~\cite{dvc} proposed the first end-to-end video compression framework called DVC, which uses optical flow for inter-prediction and replaces the DCT transform with an autoencoder. 
Lu \emph{et al.}~\cite{fvc} proposed FVC to convert videos from the pixel domain to the feature domain and use deformable convolution for motion estimation and motion compensation in the feature domain. In traditional hybrid coding frameworks and above neural video compression frameworks, residuals are calculated based on mathematical subtraction. This method is simple and easy to implement, but it may not be the optimal solution for compression. Li \emph{et al.}~\cite{dcvc} redefined the concept of residual and transform subtraction-based residual into conditional residual calculated by the neural codec, named DCVC. Sheng \emph{et al.}~\cite{dcvc_tcm} proposed the DCVC-TCM with a multi-scale conditional residual, which enhances the ability to remove inter-frame temporal redundancy. Overall, the existing neural video compression methods achieve improved compression efficiency from various perspectives such as inter-prediction, residual compression, and entropy models. Many NVC methods (e.g., the DCVC series~\cite{dcvc_dc, dcvc_hem, dcvc_fm}) demonstrate formidable compression capabilities.

\subsection{Neural-based Video Coding for Machines}

The exploration of neural-based VCM reveals two pivotal paradigms: the compress-then-analyze (CTA) paradigm and the analyze-then-compress (ATC) paradigm.

\subsubsection{Image and Video Compression in the CTA Paradigm}

With the surge in machine vision applications, video compression frameworks are re-envisioned to better cater to downstream machine vision tasks. Some methods~\cite{cta_joint, cta_semanticprior} bridge the image codec and downstream task networks, and then integrate the loss function of the downstream task to guide the optimization process of the compression network, thus tailor-fitting it for attaining enhanced performance on specific tasks. 
In addition, Tian \emph{et al.}~\cite{smc++, smc} proposed maintaining semantic similarities through an additional bitstream, which improves the performance on multiple downstream tasks in an unsupervised way. 
Furthermore, the introduction of plug-and-play preprocessing modules~\cite{cta_preprocess3} represents a significant improvement. These approaches achieve better downstream performance by enhancing important regions and filtering useless details for downstream analysis. 
Moreover, VCM is a sub-task of video coding for humans and machines~(VCHM). Some studies~\cite{scalable_choi, scalable_dcc} proposed to modify the decoder and use features that are originally dedicated to fully reconstructing images for downstream video analysis.

\subsubsection{Feature Compression in ATC Paradigm}

Intermediate feature compression is a widely studied VCM method under the ATC paradigm. 
Intermediate features contain more general information about images than high-level features do and offer the potential for conducting multitask analysis. Moreover, they preserve the original spatial structure, which enables more effective redundancy removal through neural networks. Unlike shallow features, intermediate features undergo a preliminary extraction process, where irrelevant information is filtered out for machine vision tasks, making it easier to compress.
In image feature compression, some approaches adopt traditional hybrid codec~\cite{video_codec_chen2019lossy} or variational autoencoder (VAE)-based networks that are optimized by feature distortion and specific task losses~\cite{end2endFeatureCompression2} for intra-compression.
Moreover, some methods~\cite{end2endFeatureCompression_multiscale1, end2endFeatureCompression_multiscale2} change the compressed object from a single intermediate feature to multi-scale features and compress them into a joint bit stream. 
In the field of video features compression, Misra~\emph{et al.}~\cite{misra2022video} introduced an end-to-end feature compression network. It employs a simple ResBlock-based~\cite{resnet50} bidirectional interpolation in the feature domain, and the entire framework is optimized for specific downstream tasks. Sheng~\emph{et al.}~\cite{sheng2023lvvc} proposed a framework that conducts pixel-feature-domain inter-compression and supports multiple downstream tasks by freezing the upstream codec and optimizing the downstream networks. However, a limitation is encountered when retraining and redeploying the upstream feature codec or the whole downstream task networks in practical applications. 
In light of the above challenge, there is a growing need for adaptable and scalable VCM solutions. 

\begin{figure}[!t]
    \centering
    \includegraphics[width=\linewidth]{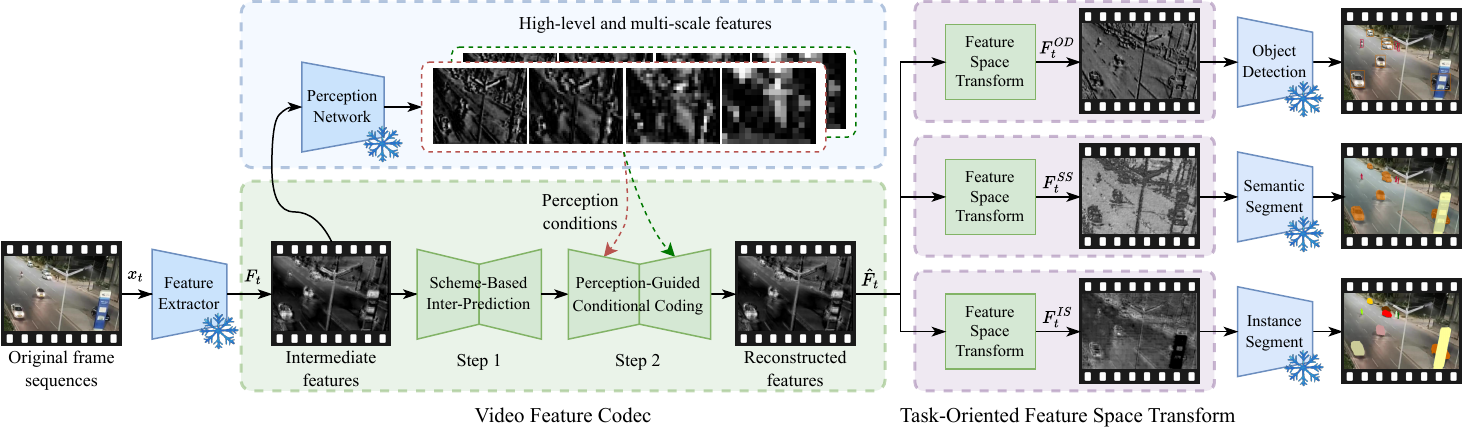}
    \caption{Overview of the proposed TransVFC framework. $F_t^{OD}$, $F_t^{SS}$, and $F_t^{IS}$ denote features for downstream object detection, semantic segmentation, and instance segmentation tasks, respectively. The TransVFC framework employs a compress-then-transfer process. Specifically, the codec conducts scheme-based inter-prediction to form a coarse compensated feature and then performs perception-guided conditional coding for fine reconstruction. The reconstructed features are subsequently transferred to other feature spaces of diverse downstream machine vision tasks via the FST modules. Notably, each downstream task corresponds to a distinct FST module. }
    \label{fig_overview}
\end{figure}

\section{Methodology}

The pipeline of the proposed Transformable Video Feature Compression (TransVFC) framework is shown in Figure~\ref{fig_overview}. It contains two main components: a neural-based video feature codec and diverse feature space transform (FST) modules. 
Inspired by \cite{misra2022video, fvc}, the intermediate features are extracted by the $res2$ layers of the ResNet50 backbone in Faster R-CNN~\cite{faster-rcnn}; then, the 256D features are converted into a 64D representation to squeeze their channel redundancy.  
The video feature codec follows the motion-then-residual paradigm; it employs the scheme-based inter-prediction module to obtain a coarse motion-compensated feature and then uses the perception-guided conditional coding module for fine feature reconstruction.  
Afterward, the FST modules transfer the reconstructed intermediate feature $\hat{F_t}$ to different feature spaces, making them suitable for various downstream machine vision tasks. Notably, each downstream task is associated with a dedicated FST module.

\begin{figure}[!t]
\centering
\includegraphics[width=0.9\linewidth]{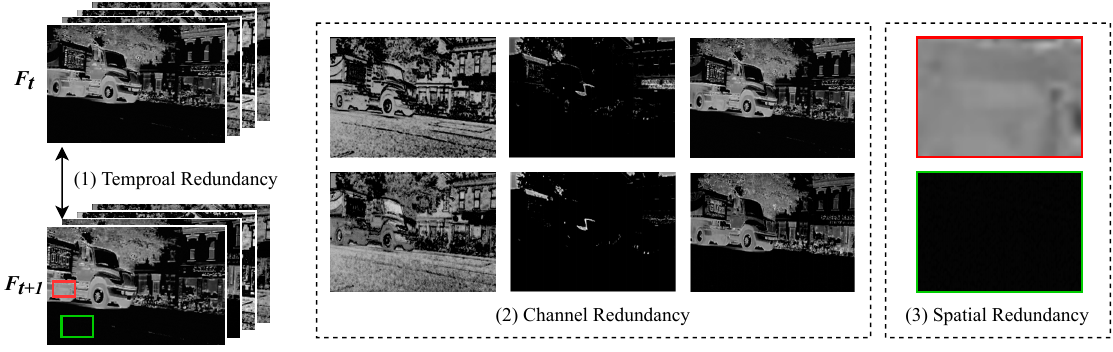}
\caption{Temporal, channel, and spatial redundancies among video features. The above redundancies need to be squeezed by the video feature codec. }
\label{fig_intermediate_channel}   
\end{figure}

\begin{figure}[!h]
    \centering
    \includegraphics[width=\linewidth]{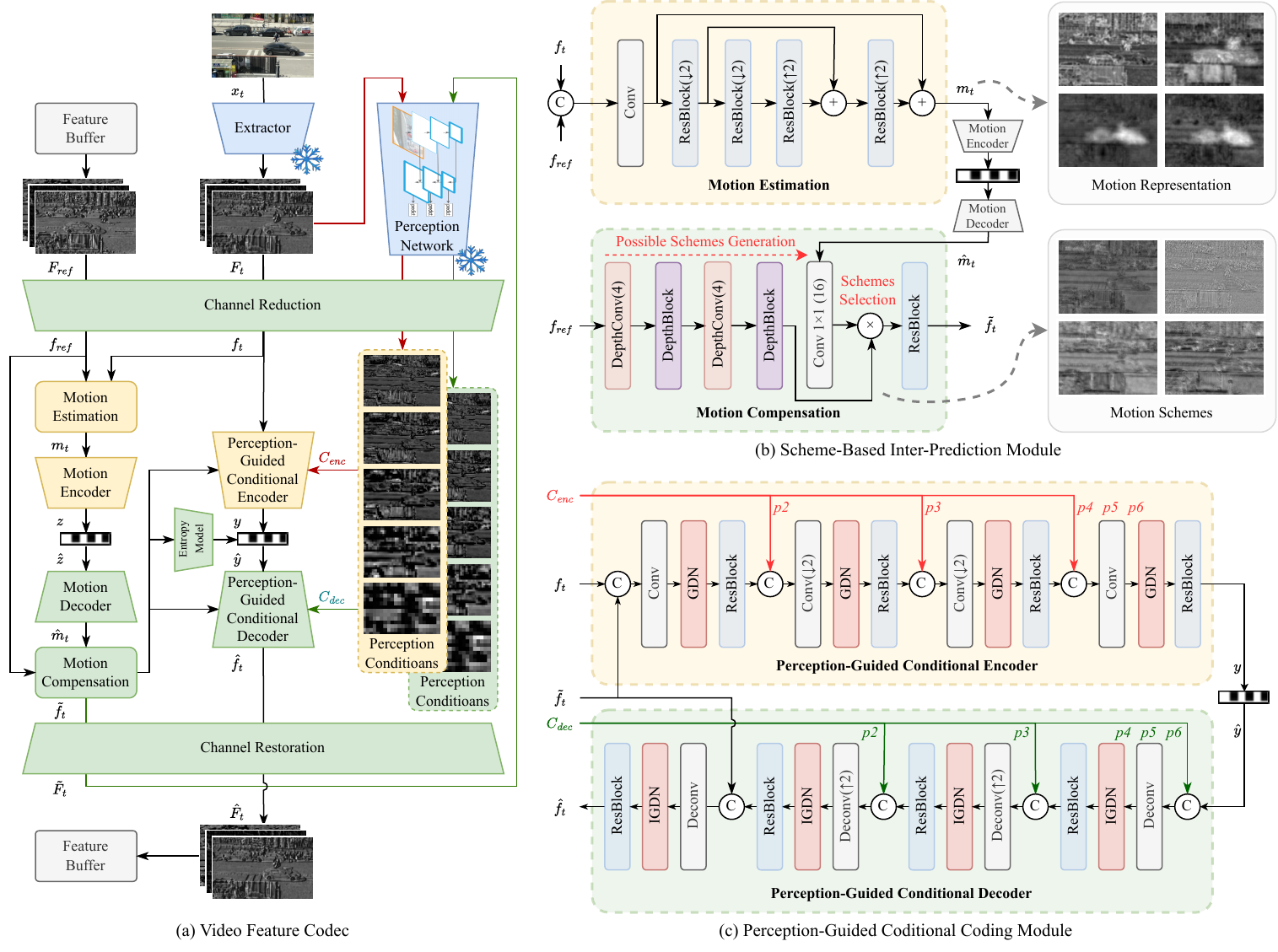}
    \caption{(a) The overall structure of the proposed neural-based video feature codec. It contains 3 main stages: channel reduction/restoration, scheme-based inter-prediction, and perception-guided conditional coding. The green modules are located on both the encoder and decoder sides, whereas the yellow modules are only used on the encoder side. 
    (b) The structure of the scheme-based inter-prediction module, including a motion estimation module, a motion compensation module, a motion encoder, and a motion decoder. $DepthConv(n)$ represents a depthwise separable convolution layer with the number of channels increased by $n$ times. The structure of $DepthBlock$ is similar to $ResBlock$ but replaces the convolution layers with the depthwise separable convolution layers.
    (c) The structure of the perception-guided conditional encoder and decoder. High-level and multi-scale features $C_{enc}$ and $C_{dec}$ are inferred from the Perception Network and used as conditions during the residual compression and reconstruction phases, respectively.}
    \label{graph_feature_codec}
\end{figure}

\subsection{Scheme-Based Inter-Prediction}

Temporal redundancy exists among repeated spatial structures, thus highlighting the need for redundancy removal by inter-prediction techniques, as shown in Figure~\ref{fig_intermediate_channel}. For conducting inter-prediction in the feature domain, the deformable-convolution-based approach~\cite{fvc} focuses on finding the optimal reference region and recombining existing feature values. To better address complex motion, we depart from this referencing-and-recombination method. Instead, we propose a scheme-based inter-prediction module. It generates a variety of potential motion schemes from the reference frame and selectively combines them to obtain the compensated feature.

In the encoder, the motion estimation module performs a four-step sampling procedure for motion analysis across three distinct scales. The motion representation $m_t$ contains both the global trends and the high-frequency details of motion, as shown in Figure~\ref{graph_feature_codec}(b). Afterward, the motion encoder compresses $m_t$ into a compact latent representation $z$ with dimensions of $(H/16, W/16, 64)$. Subsequently, the latent representation $z$ is quantized into $\hat{z}$ for entropy coding and transmission. 
In the decoder, the motion combination matrix $\hat{m}_t$ is reconstructed by the motion decoder module from $\hat{z}$. Leveraging the channel-wise computation by the depthwise separable convolution~\cite{chollet2017xception}, the motion compensation module generates diverse possible motion schemes based on the reference frame $f_{ref}$. Then, referring to the motion representation $\hat{m}_t$, schemes are judiciously selected and combined to form the compensated feature $\Tilde{f}_t$.
More analyses and visualizations are shown in Section IV.
The whole scheme-based inter-prediction process is described in the following equation:

\begin{small}
\begin{equation}
    \Tilde{f_t} = MC(f_{ref}, \mathcal{D}_{m}(\lfloor \mathcal{E}_m(ME(f_t, f_{ref}))\rceil))
\end{equation}
\end{small}

\noindent where $ME(\cdot)$ denotes motion estimation, $MC(\cdot)$ denotes motion compensation. $\mathcal{E}_m(\cdot)$ and $\mathcal{D}_m(\cdot)$ denote motion encoder and decoder, respectively. $\lfloor \cdot \rceil$ denotes the quantization operation.

\subsection{Perception-Guided Conditional Coding}

The compensated feature $\Tilde{f_t}$ is obtained from the previous scheme-based inter-prediction. However, there is a gap in content detail between $\Tilde{f_t}$ and $f_t$, making it essential to complete the content details using the residual. We employ conditional coding to compress the residual in the feature domain. 
Since different machine vision tasks share common perceptions~\cite{x_decoder}, we further introduce multi-scale high-level features in Faster R-CNN~\cite{faster-rcnn} as perception conditions to help the reconstructed features better align with the downstream machine perception. Furthermore, the perception conditions offer TransVFC more prior knowledge during residual compression and reconstruction phases for achieving lower entropy and better spatial redundancy removal, as follows:

\begin{small}
\begin{equation}
    H(f-\Tilde{f}) > H(f|\Tilde{f}) > H(f|\Tilde{f},C_{enc},C_{dec})
\end{equation}
\end{small}

\noindent where $H(\cdot)$ represents entropy, $\Tilde{f}$ denotes the compensated feature, $C_{enc}$ and $C_{dec}$ denote the perception conditions for encoding and decoding, respectively.

As depicted in Figure~\ref{graph_feature_codec}(c), the perception-guided conditional encoder comprises a four-step feature extraction process that compresses residuals into a compact and flat representation, while the decoder mirrors this structure symmetrically to reconstruct the intermediate features. Multi-scale perception conditions are strategically inserted into positions that align with their corresponding spatial resolutions (specifically at the $1/4$, $1/8$, and $1/16$ scales), serving as conditions for both encoding and decoding to enhance the overall performance of the codec. In particular, the encoding perception conditions $C_{enc}=\{p2, p3, p4, p5, p6\}$ are inferred from the original intermediate feature $F_t$ via feature pyramid network (FPN) backbone of Faster R-CNN. Due to the invisibility of $F_t$ during decoding, the decoding perception condition $C_{dec}$ is calculated from the compensated feature $\Tilde{F_t}$. 
The whole process of perception-guided conditional coding is described as follows:

\begin{small}
\begin{equation}
    \hat{f_t} = \mathcal{D}_c(\lfloor \mathcal{E}_c(f_t | C_{enc}, \Tilde{f_t}) \rceil | C_{dec}, \Tilde{f_t})
\end{equation}
\end{small}

\noindent where $\mathcal{E}_c(\cdot)$ and $\mathcal{D}_c(\cdot)$ denote perception-guided conditional encoder and decoder, respectively.

The latent representation $y$ of the residual with dimensions of $(H/16, W/16, 96)$ is entropy-encoded by an entropy model similar to DCVC-TCM~\cite{dcvc_tcm}. With respect to computational efficiency, autoregressive or other complex techniques are not employed in TransVFC. Additionally, a detailed explanation of how the perception-guided conditional coding removes redundancy is given in Section IV.

\subsection{Task-Oriented Feature Space Transform}

\begin{figure}[!h]
\centering
\includegraphics[width=0.8\linewidth]{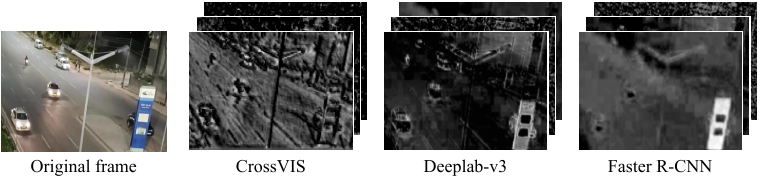}
\caption{Visualizations of the first three channels of the intermediate features across various machine vision networks. There are similar spatial structures but distinct feature patterns and textures among different intermediate features.}
\label{fig_diff}   
\end{figure}

\begin{figure}[!h]
\centering
\includegraphics[width=0.8\linewidth]{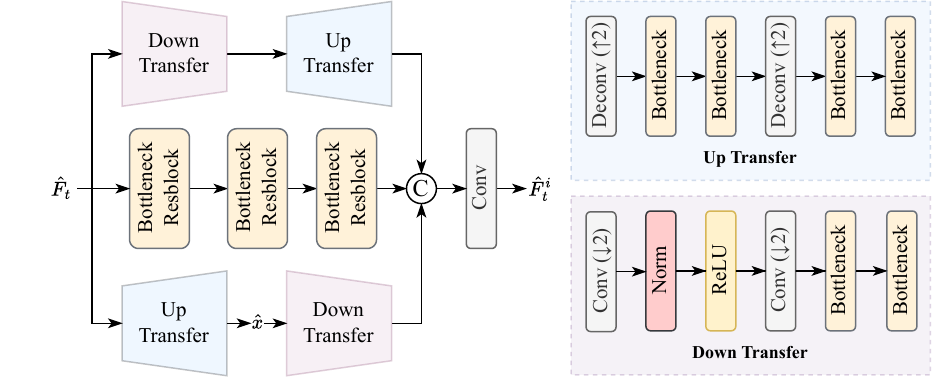}
\caption{The structure of feature space transform module. The reconstructed intermediate feature $\hat{F_t}$ is transferred to $\hat{F_t^i}$ in a new feature space for the $i$-th downstream machine vision task.}
\label{fig_fst}   
\end{figure}

Owing to the gap between the intermediate features of different neural networks, as shown in Figure~\ref{fig_diff}, the decoded video features cannot be directly used in diverse downstream tasks. Some studies \cite{scalable_choi, sheng2023lvvc} have already shown that intermediate features have the potential to be converted and used in other machine vision tasks. Inspired by~\cite{scalable_choi}, we design the multi-scale feature space transform (FST) module that maps the reconstructed features to other feature spaces for different downstream tasks. Different from the existing neural-based VCM strategies~\cite{sheng2023lvvc, misra2022video}, our approach does not fine-tune the upstream feature codec and downstream task networks. Instead, it only requires a single lightweight FST module to be trained for a specific downstream task.

As shown in Figure~\ref{fig_fst}, the FST module is structured with three branches: the up-then-down branch, which coarsely reconstructs the current frame $\hat{x_t}$ for content preservation in pixel domain; the bottleneck-resblock~\cite{resnet50} branch, facilitating feature migration at the original shape; and the down-then-up branch, focusing on global information extraction. Additionally, a convolution layer is used to align the channel and spatial shape of the output features to the specific downstream task. The process of feature space transform is described below:

\begin{small}
\begin{equation}
\hat{F_t^i} = FST^i(\hat{F_t})
\end{equation}
\end{small}

\noindent where $FST^i(\cdot)$ denotes the $i$-th FST module, $\hat{F_t}$ denotes the intermediate feature reconstructed by the video feature codec, and $\hat{F_t^i}$ denotes the transferred feature that is suitable for the $i$-th downstream task.

\subsection{Optimization}

\begin{figure}[!h]
\centering
\includegraphics[width=0.75\linewidth]{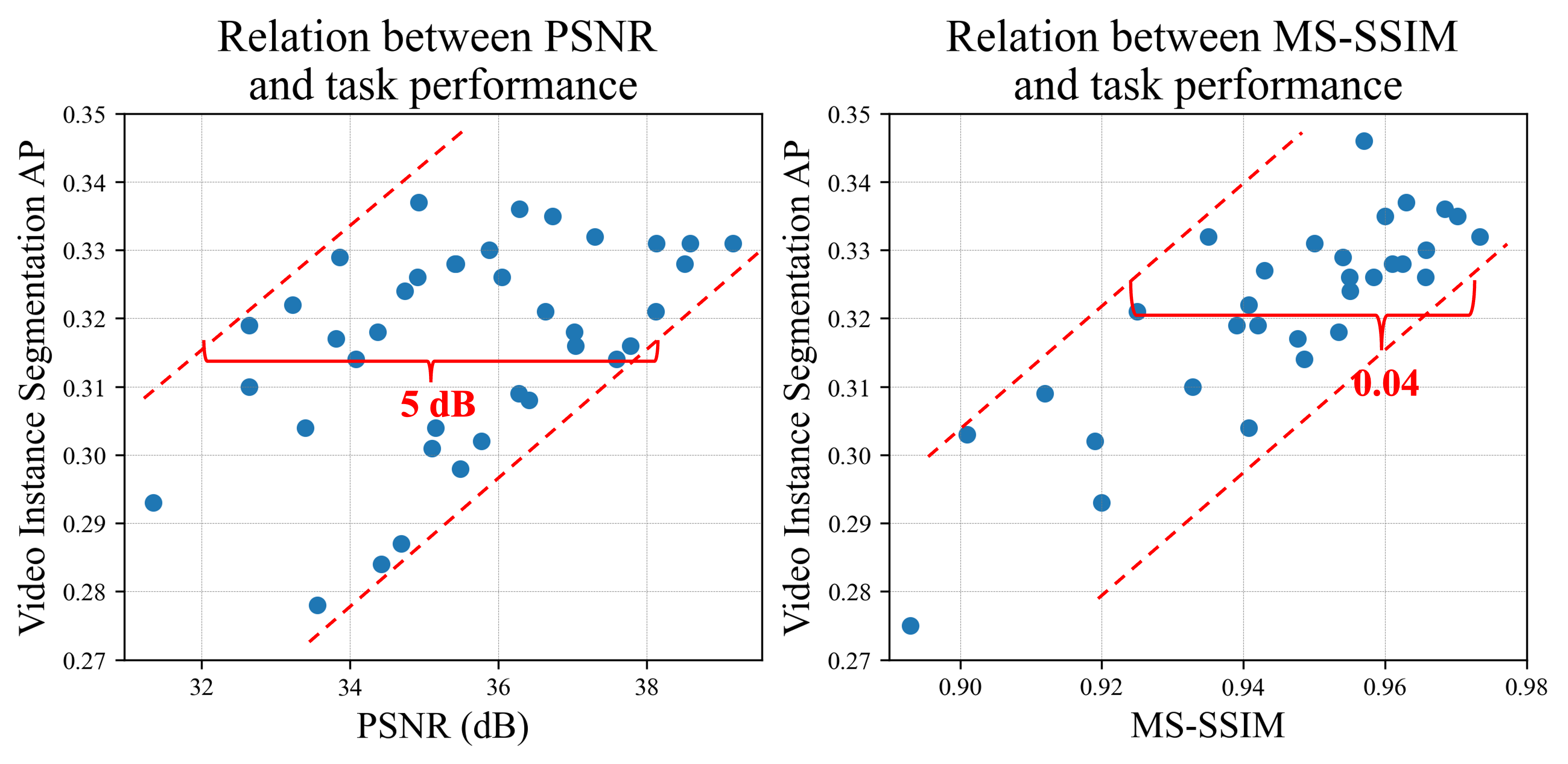}
\caption{The correlations between the pixel-domain and HVS-related distortion metric (the PSNR and MS-SSIM) and the downstream machine vision performance (e.g., the average precision of video instance segmentation) are weak. 
Optimizing the compression network for minimizing pixel-domain HVS distortions is not the best approach for the VCM scenario. The degraded videos are collected from traditional hybrid codecs and neural-based codecs~\cite{h265, dcvc_dc, dcvc_hem, dcvc_tcm, dcvc, fvc}. Particularly, the Youtube-VIS 2019 dataset and the CrossVIS~\cite{crossvis} model are used.}
\label{fig_hvs_vcm}   
\end{figure}

Since strong correlations between HVS-oriented pixel-domain metrics and machine vision performance are lacking, as mentioned in Figure~\ref{fig_hvs_vcm}. The optimization of our proposed TransVFC is mainly conducted in the feature domain and divided into two stages. 

\subsubsection{Optimization of the Video Feature Codec}

Rate-distortion optimization (RDO) is performed for the proposed video feature codec in the feature domain. The loss function $\mathcal{L}_{codec}$ is defined as follows: 

\begin{small}
\begin{equation}
    \mathcal{L}_{codec} = \lambda_R (R_{r} + R_{m}) + \lambda_{f} D_{f} + \lambda_{c} D_{c} + \lambda_{p} D_{p} 
\end{equation}
\end{small}

\noindent where $\lambda_R$, $\lambda_{f}$, $\lambda_{c}$, and $\lambda_{p}$ represent coefficients for balancing different loss terms. $R_{r}$ and $R_{m}$ represent the bitrates of the residual and motion representation, respectively. $D_{f}$ denotes the mean square error (MSE) between the original intermediate feature $F_t$ and the reconstructed feature $\hat{F_t}$. $D_{c}$ denotes the MSE between $F_t$ and compensated feature $\Tilde{F_t}$. $D_{p}$ denotes the distortion in the perception space and is defined as follows:

\begin{small}
\begin{equation}
    D_{p} = \frac{1}{N} \sum_{j=1}^{N=5} MSE(PN_j(F_t), PN_j(\hat{F_t}))
\end{equation}
\end{small}

\noindent where $PN(\cdot)$ denotes the perception network contained in TransVFC, and $N$ denotes the number of high-level output features derived from the perception network. 

\subsubsection{Optimization of Feature Space Transform Module}

Since the FST module mainly transforms the reconstructed intermediate features to other spaces for downstream networks. It is optimized for minimizing the downstream task loss and feature distortion in the new feature space. All other modules are frozen in this training stage. The total loss of the FST modules is defined as follows:

\begin{small}
\begin{equation}
    \mathcal{L}_{FST} = \lambda_{task} \mathcal{L}_{task} + \lambda_x D_x + \lambda_{mid} D_{mid} + \lambda_{high} D_{high}
\end{equation}
\end{small}

\noindent where $\lambda_{task}$, $\lambda_{x}$, $\lambda_{mid}$ and $\lambda_{high}$ represent coefficients for balancing different loss terms. $\mathcal{L}_{task}$ denotes the loss of downstream task network. $D_{mid}$ denotes the MSE between the transferred feature $\hat{F_t^i}$ and the original feature $F_t^i$ for the $i$-th downstream task, $D_x$ denotes the MSE between the coarsely reconstructed frame $\hat{x_t}$ and the original frame $x_t$, and the definition of $D_{high}$ is defined as follows:

\begin{small}
\begin{equation}
    D_{high} = \frac{1}{N} \sum_{j=1}^{N} MSE(TASK^i_j(F_t^i), TASK^i_j(\hat{F_t^i}))
\end{equation}
\end{small}

\noindent where $TASK^i(\cdot)$ represents the backbone of the $i$-th downstream task network and $N$ denotes the number of output high-level features from the $i$-th downstream backbone.

\section{Experiment}

\subsection{Experimental Settings}

\subsubsection{Downstream Machine Vision Tasks}

The performance of TransVFC is verified on three downstream tasks at different granularities. We employ the CrossVIS~\cite{crossvis} framework for video instance segmentation, DeepLab-v3~\cite{deeplabv3} for semantic segmentation, and Faster R-CNN~\cite{faster-rcnn} for object detection. The parameters of all the downstream networks are frozen throughout the experiments. 

\subsubsection{Datasets}
Experiments are conducted on the YoutubeVIS2019 (YTVIS2019)\cite{crossvis} and Video Scene Parsing in the Wild (VSPW)~\cite{vspw} datasets. The YTVIS2019 dataset is a large video dataset that includes 2,883 videos with frame-level annotations of 40 categories for video instance segmentation. The resolutions of the videos range from 1080P to 360P, and the data preproces stages follow \cite{crossvis}. The VSPW dataset is a large video dataset that includes 3,536 videos in 480P resolution across 231 scenarios. It has frame-level annotations of 124 categories for video semantic segmentation. 

\subsubsection{Compared Methods}

The proposed TransVFC framework is compared with traditional hybrid codecs VTM-23.1 (lowdelay-P)~\cite{vtm}, HM-18.0 (lowdelay-P)~\footnote{The command of VTM and HM is \texttt{./bin/TAppEncoderStatic -c \\ ./cfg/encoder\_lowdelay\_P\_main.cfg -i \{input\_path\} -b \{output\_binary\_path\} -o \{output\_path\} -wdt \{width\} -hgt \{height\} -q \{QP\} -fr \{frame\_rate\} -InputChromaFormat=420 --IntraPeriod=12}}~\cite{hm} and x265 (FFmpeg-4.2.7, zerolatency)~\footnote{The command of x265 is \texttt{FFREPORT=file=ffreport.log:level=56 ffmpeg -pix\_fmt yuv420p -s \{width\}x\{height\} -i \{input\_path\} -c:v libx265 -tune zerolatency -x265-params "crf=\{crf\}:keyint=12:verbose=1" out.mkv}}~\cite{x265}, and open-sourced neural video compression (NVC) frameworks, such as DCVC-DC~\cite{dcvc_dc}, DCVC-HEM~\cite{dcvc_hem}, DCVC-TCM~\cite{dcvc_tcm}, DCVC~\cite{dcvc}, and FVC~\cite{fvc}. For compared NVC methods, all available pre-trained models are evaluated across different metrics (PSNR, MS-SSIM, YUV), showcasing only the model with the highest rate-task performance. In addition, VCM-oriented video codec SMC++~\cite{smc++} is used as a comparison method. 

\subsubsection{Implementation Details}

\begin{table}[!h]
\centering
\scriptsize \footnotesize
\setlength\tabcolsep{9pt} 
\begin{tabular}{ccc}
\hline
Stages & $\mathcal{L}_{codec}$ & Learning rate\\
\hline
1 & $\frac{1}{2} \lambda_R R_y + \lambda_{f} D_{f} + \lambda_{c} D_{c} + \lambda_{p} D_{p} $ & $1 \times 10^{-4}$\\
2 & $\frac{1}{2} \lambda_R (R_y + R_z) + \lambda_{f} D_{f} + \lambda_{c} D_{c} + \lambda_{p} D_{p} $ & $1 \times 10^{-4}$\\
3 & $\lambda_R (R_y + R_z) + \lambda_{f} D_{f} + \lambda_{c} D_{c} + \lambda_{p} D_{p} $ & $1 \times 10^{-4}$\\
4 & $\lambda_R (R_y + R_z) + \lambda_{f} D_{f} + \lambda_{c} D_{c} + \lambda_{p} D_{p} $ & $5 \times 10^{-5}$\\
5 & $\lambda_R (R_y + R_z) + \lambda_{f} D_{f} + \lambda_{c} D_{c} + \lambda_{p} D_{p} $ & $1 \times 10^{-5}$\\
\hline
\end{tabular}
\caption{Training strategy for video feature codec}
\label{table_traning_details}
\end{table}

\begin{table}[!h]
\centering
\scriptsize \footnotesize
\setlength\tabcolsep{10pt} 
\begin{tabular}{ccccc}
\hline
Downstream tasks &$\lambda_{mid}$  & $\lambda_{high}$ & $\lambda_{x}$ & $\lambda_{task}$\\
\hline
Object detection & 16 & 4 & 1024 & 10 \\
Instance segmentation & 8 & 64 & 1024 & 1 \\
Semantic segmentation & 16 & 64 & 1024 & 10 \\
\hline
\end{tabular}
\caption{Training hyperparameters $\lambda$ for feature space transform module}
\label{table_traning_details_2}
\end{table}

In the first stage, we optimize the video feature compression framework at different bitrates with $\lambda_{R} = 16, 32, 128, 256$, $\lambda_{f}=16$, $\lambda_{c}=0.1\lambda_{f}$, and $\lambda_{p}=4$. The training strategy is shown in Table~\ref{table_traning_details}. The input features during training are cropped to $128 \times 128$. The neural-based video feature codec is optimized on the YTVIS2019-train.

In the second stage, different weights $\lambda$ are used to train FST modules for each downstream task, as shown in Table~\ref{table_traning_details_2}. 
\textcolor{\revisioncolor}{Owing to the impracticality of exhaustively tuning the $\lambda$ weights under computational constraints, we determine them empirically. Inspired by the existing approach~\cite{deeplabv3}, we follow a simple rule: each loss term is scaled such that its magnitude and the gradient it contributes to the FST module are comparable to those of others terms. This strategy facilitates stable training and balanced learning across multiple objectives.} The number of training iteration for the FST module is $100k$, and the learning rate is set to $1 \times 10^{-5}$. \textcolor{\revisioncolor}{Notably, the FST modules for different downstream tasks are trained separately. }

The implementation of TransVFC is based on PyTorch 1.9.0. The whole framework is optimized on a single NVIDIA RTX 3090 24GB with $batchsize=4$.

\textcolor{\revisioncolor}{\subsection{Evaluation Metrics}}

\textcolor{\revisioncolor}{The number of bits per pixel (bpp) is used to represent bitrate cost, where a lower bpp value indicates a higher compression ratio. For downstream tasks, average precision (AP)~$\uparrow$ is used to evaluate the performance of object detection and instance segmentation, referring to~\cite{faster-rcnn, crossvis}. While the mean intersection over union (mIoU)~$\uparrow$ is used to assess semantic segmentation performance, referring to~\cite{deeplabv3}. The PSNR~$\uparrow$ and MS-SSIM~$\uparrow$ are employed to evaluate the quality of the reconstructed frames. The Bjøntegaard Delta Rate (BD-Rate)~$\downarrow$ is used to quantify the overall rate-task performance. It reflects the percentage of bitrate change while achieving the same task performance. A lower BD-Rate represents greater bitrate savings. VTM-23.1 serves as the anchor for calculating the BD-Rate. }

\begin{table}[!h]
\centering
\scriptsize \footnotesize
\begin{tabular}{lccc}
\hline
                                    & Object            & Semantic          & Instance \\
                                    & detection         & segmentation      & segmentation \\
\hline
VTM-23.1 (low-delay)~\cite{vtm}               & 0.00              & 0.00              & 0.00 \\
HM-18.0 (low-delay)~\cite{hm}                 & 7.82              & \textbf{-11.40}   & 5.60 \\
x265 (zero-latency)~\cite{x265}                & -1.16             & 36.34             & 3.91 \\
\hline
FVC (CVPR'20)~\cite{fvc}             & 97.15             & 130.03            & 368.56 \\
DCVC (NerulPS'21)~\cite{dcvc}        & 50.34             & 286.38            & 109.43 \\
DCVC-TCM (TMM'22)~\cite{dcvc_tcm}    & 7.84              & 204.46            & 32.69 \\
DCVC-HEM (ACMMM'22)~\cite{dcvc_hem}  & -3.92             & 183.80            & 46.34 \\
DCVC-DC (CVPR'23)~\cite{dcvc_dc}     & -4.53             & 145.53            & 26.56 \\
\hline
SMC++ (arXiv'24)~\cite{smc++} & -4.28 & 74.61 & -6.77 \\
TransVFC (Ours)                  & \textbf{-15.21}   & 63.60             & \textbf{-27.67} \\
\hline
\end{tabular}
\caption{BD-Rate (\%) $\downarrow$ comparison. The anchor is VTM-23.1. \textbf{Bold} indicates the best results.}
\label{table_bd_rate}
\end{table}

\begin{figure}[!h]
    \centering
    \includegraphics[width=\linewidth]{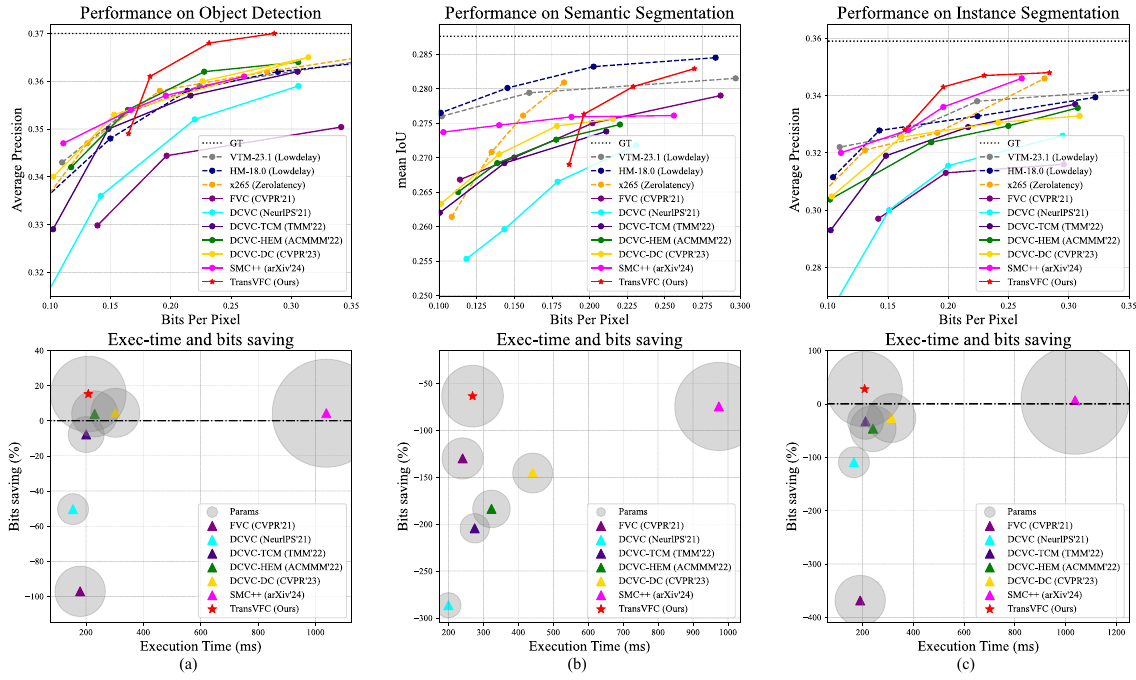}
    \caption{Rate-task performance of all the compared methods (in the upper row) and execution times of neural-based methods (in the lower row) on object detection, semantic segmentation, and instance segmentation tasks. The execution time, including compression and downstream analysis, is evaluated with $batchsize=1$ on a single NVIDIA RTX 3090 24GB, excluding the time of file I/O. }
    \label{curve-whole}
\end{figure}

\subsection{Rate-Task Performance}

\subsubsection{Object Detection}

\begin{figure}[t]
    \centering
    \includegraphics[width=0.95\linewidth]{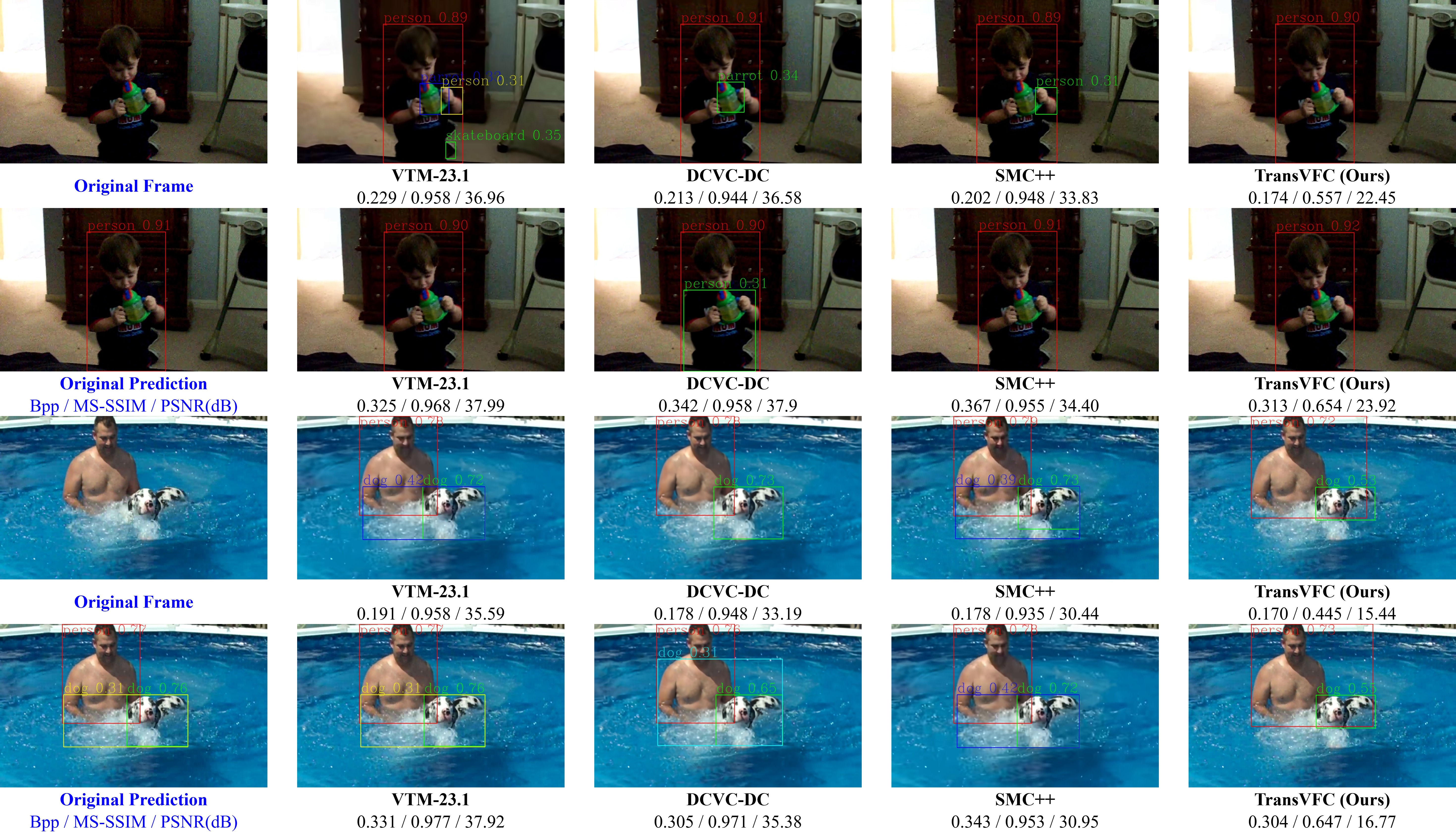}
    \caption{\textcolor{\revisioncolor}{Visualization of object detection results at different bitrates, along with the corresponding bpp, MS-SSIM, and PSNR values. The proposed TransVFC tends to produce fewer false detections. In contrast, other methods exhibit false positive detections, despite achieving high reconstruction quality. }}
    \label{figure-visualization-od}
\end{figure}

The implementation of the Faster R-CNN~\cite{faster-rcnn} is based on Detectron2~\cite{wu2019detectron2}, which is an extensively used and efficient framework for keypoint detection, object detection, and segmentation.  
The task performance across various bitrates is displayed in Figure~\ref{curve-whole}(a). In terms of rate-task performance, TransVFC achieves a 15.21\% bitrate reduction relative to VTM, as shown in Table~\ref{table_bd_rate}. From a rate-time perspective, TransVFC achieves a better speed-performance balance than the other neural-based methods do. \textcolor{\revisioncolor}{Visualization examples of the object detection results are shown in Figure~\ref{figure-visualization-od}. TransVFC tends to produce fewer false detections under various bitrates and scenarios. }

\subsubsection{Semantic Segmentation}

\begin{figure}[t]
    \centering
    \includegraphics[width=0.95\linewidth]{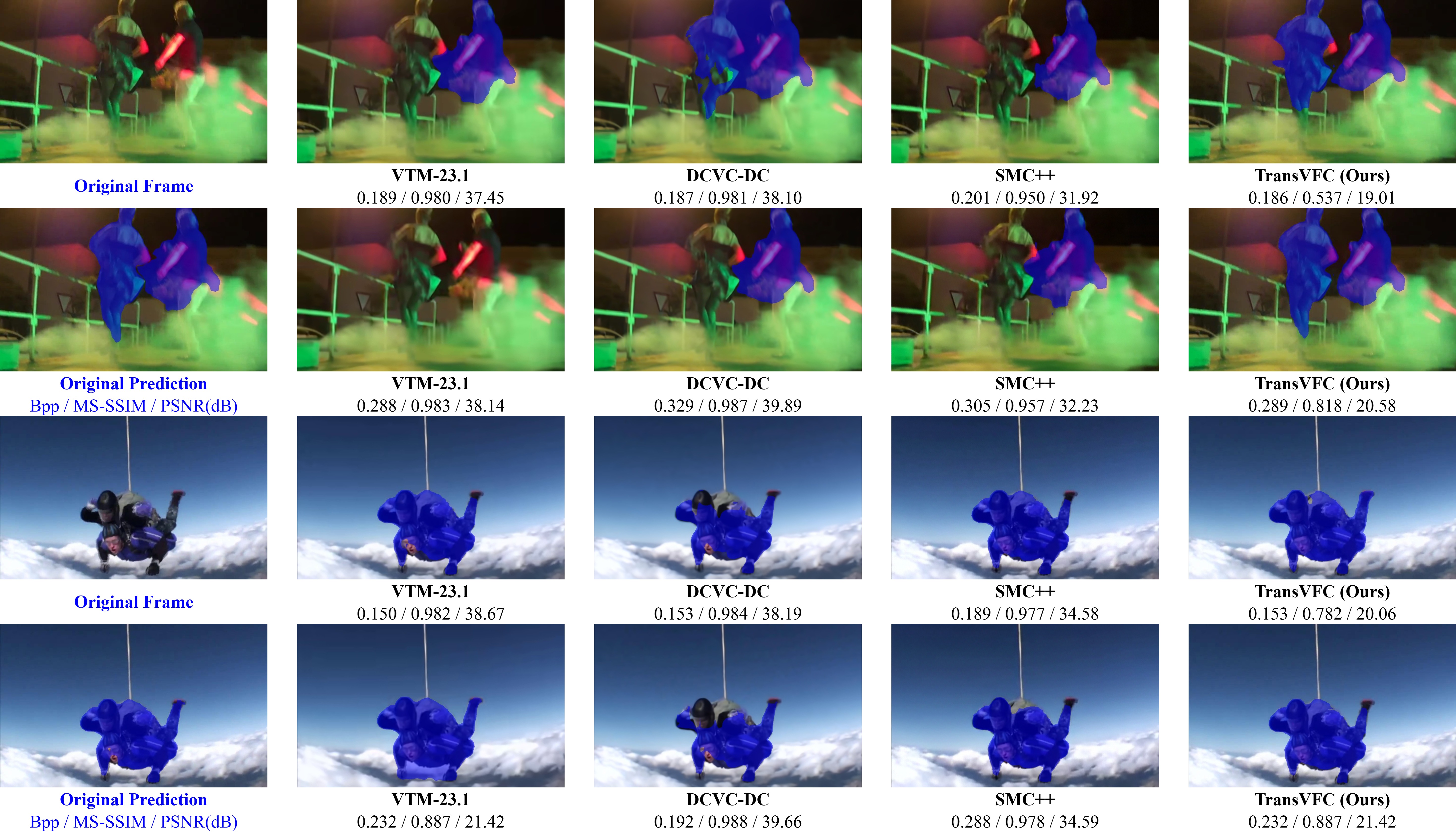}
    \caption{\textcolor{\revisioncolor}{Visualization of semantic segmentation, bpp, MS-SSIM, and PSNR at different bitrates. TransVFC better preserves object contours and performs segmentation more accurately under challenging conditions (e.g., fog, low-light, and intense motion) compared to other methods.}}
    \label{figure-visualization-ss}
\end{figure}

We implement the DeepLab-v3~\cite{deeplabv3} via TorchVision-0.9.0. Following~\cite{deeplabv3}, mean intersection over Union (mIoU) is used to evaluate the semantic segmentation performance of the tested methods. As demonstrated in Table~\ref{table_bd_rate}, TransVFC outperforms the best neural-based method, SMC++~\cite{smc++}, in terms of rate-task performance. Additionally, TransVFC achieves the best speed-performance balance among the tested neural-based methods, as shown in Figure~\ref{curve-whole}(b). \textcolor{\revisioncolor}{Visualization of semantic segmentation is provided in Figure~\ref{figure-visualization-ss}, where TransVFC demonstrates better preservation of object contours and more accurate segmentation under challenging conditions, such as fog and low-light, compared to other methods. }

\begin{figure}[!h]
    \centering
    \includegraphics[width=0.95\linewidth]{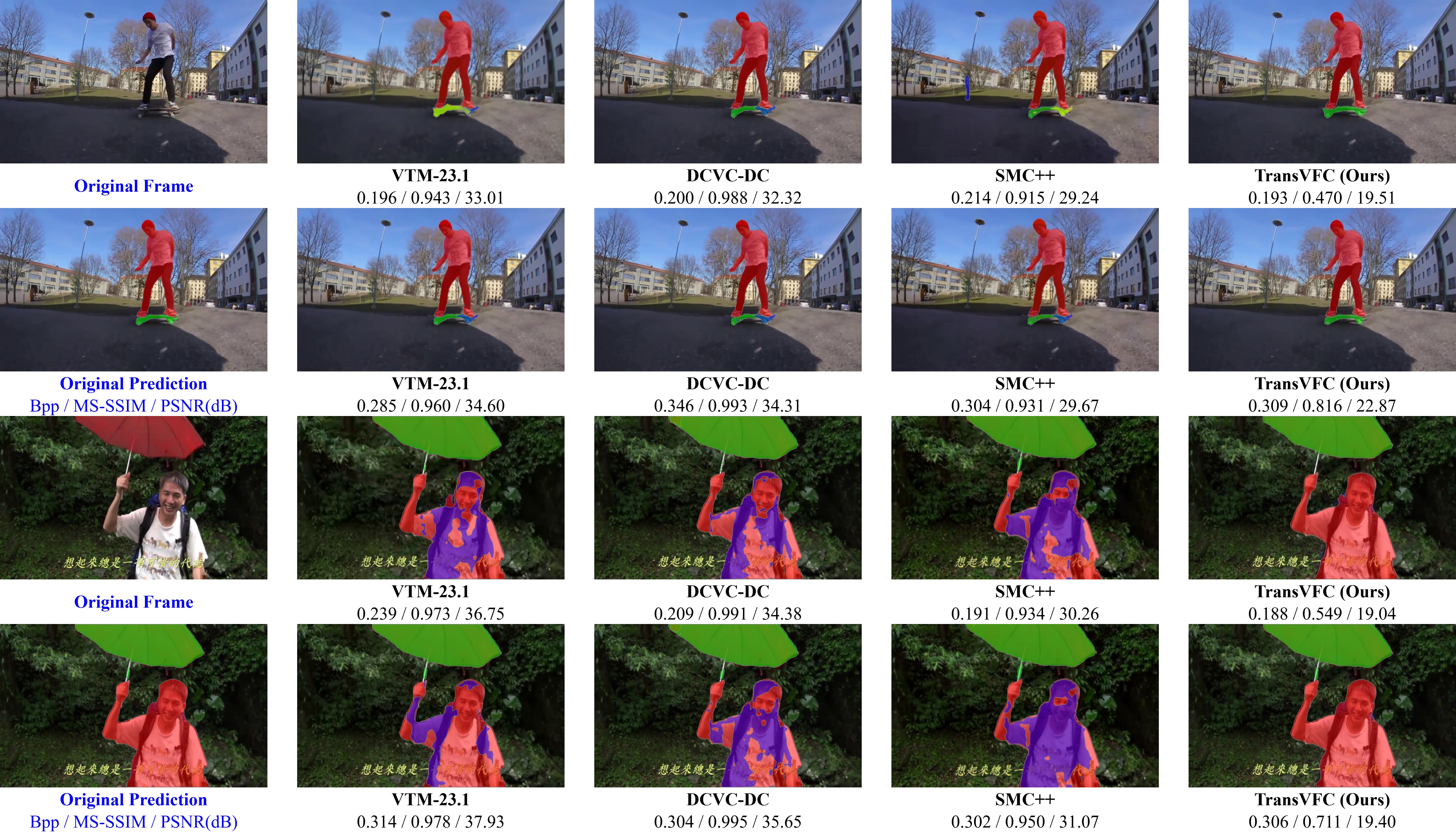}
    \caption{\textcolor{\revisioncolor}{Visualization of video instance segmentation, bpp, MS-SSIM, and PSNR at different bitrates. In videos with intense motion (in the first two rows) and tiny movement (in the last two rows), TransVFC maintains the original prediction and ensures consistency of the instance. The visualization shows that even if the reconstructed frames have high reconstruction quality, they may still underperform in downstream tasks. }}
    \label{figure-visualization-vis}
\end{figure}

\subsubsection{Instance Segmentation} 

CrossVIS~\cite{crossvis} is implemented on its officially released code. As demonstrated in Table~\ref{table_bd_rate}, in terms of rate-task performance, TransVFC achieves the highest compression ratio, achieving a 27.67\% bitrate reduction compared to VTM. In terms of execution speed, compared with the high-performance NVC method DCVC-DC~\cite{dcvc_dc}, TransVFC has a 34\% faster execution speed, as shown in Figure~\ref{curve-whole}(c). As shown in Figure~\ref{figure-visualization-vis}, TransVFC produces better subjective segmentation results at different bitrates. Despite the high quality of reconstructed frames, the downstream task network CrossVIS struggles with maintaining the segmentation consistency of the main objects (e.g., the skateboard and the man holding an umbrella), often incorrectly segmenting them into multiple instances. In contrast, our framework better maintains the consistency of the instance and maximally retains the original segmentation results.

\subsection{Analysis}

\subsubsection{Complexity of Video Features Compression}

\begin{table}[!h]
\centering
\scriptsize \footnotesize
\setlength\tabcolsep{2pt}
\begin{tabular}{lccccc}
\hline
                                    & Non-stream    & \multicolumn{2}{c}{With bitstream} & Model & MACs per\\
\cline{3-4}
                                     & inference     & Encoding        & Decoding       & params    & pixel\\
\hline
FVC (CVPR'20)~\cite{fvc}             & 165.8        & /             & /                & 21.0M      & / \\
DCVC (NerulPS'21)~\cite{dcvc}        & 129.1        & 1818.9        & 4738.3           & 7.9M       & 1.09M \\
DCVC-TCM (TMM'22)~\cite{dcvc_tcm}    & 197.6        & 232.8         & 121.4            & 10.7M      & 1.40M \\
DCVC-HEM (ACMMM'22)~\cite{dcvc_hem}  & 240.6        & 250.3         & 124.6            & 17.5M      & 1.58M \\
DCVC-DC (CVPR'23)~\cite{dcvc_dc}     & 347.5        & 285.5         & 243.8            & 19.8M      & 1.27M \\
SMC++ (arXiv'24)~\cite{smc++}      & 830.1        & /             & /                  & 96.2M      & \\
TransVFC (Ours)                      & 191.2       & 234.5         & 122.5             & 22.4+26.7M & 1.16M \\
\hline
\end{tabular}
\caption{Execution time (ms) on 720P frame, number of model parameters, and MACs per pixel of neural-based methods. }
\label{table_time}
\end{table}

We compare the execution time, number of parameters, and MACs of our proposed video feature codec with other neural-based compression methods~\cite{fvc, dcvc, dcvc_tcm, dcvc_hem, dcvc_dc, smc++}, as shown in Table~\ref{table_time}. Our proposed video feature codec consists of an optimized codec with 22.4M parameters and a frozen perception network with 26.7M parameters. The inference time reflects the computational complexity of all the neural-based modules on the GPU without including arithmetic coding. The encoding and decoding times include the arithmetic coding operation time but exclude file I/O time. \textcolor{black}{Although our codec has more parameters than other neural compression methods, it has fewer MACs per pixel than high-performance codecs like DCVC-DC and DCVC-HEM. TransVFC has a better complexity-performance balance than other high-performance NVC approaches, with efficiency gains stemming from three aspects.} First, the intermediate features have a 1/4 spatial size of the original image, which helps TransVFC use fewer convolutions and down/upsampling operations than neural video compression frameworks. Second, to improve encoding and decoding speed, TransVFC uses a simple entropy model including a mean-scale hyperprior module and a temporal prior module~\cite{dcvc_tcm}, which is better parallelized. \textcolor{black}{Third, introducing depthwise convolution reduces the computational complexity of the model~\cite{dcvc_dc, chollet2017xception}, resulting in lower MACs. }

\subsubsection{Complexity of Feature Space Transform}

The FST module in the TransVFC framework is lightweight, with a parameter size of 4.3M. It is significantly smaller than the networks used for downstream visual analysis (e.g., the CrossVIS-ResNet50 version has 37.4 M parameters), adding less additional training overhead. The execution time of FST under a 720P resolution is 11.7 ms, which accounts for only 3.3\% of the total time. \textcolor{black}{The MACs per pixel of the FST module are 0.16M.} The above results indicate that the FST module is highly effective during both the training and inference stages. 

\subsubsection{Visualization of the Scheme-Based Inter-Prediction Module}

\begin{figure}[!h]
\centering
\includegraphics[width=\linewidth]{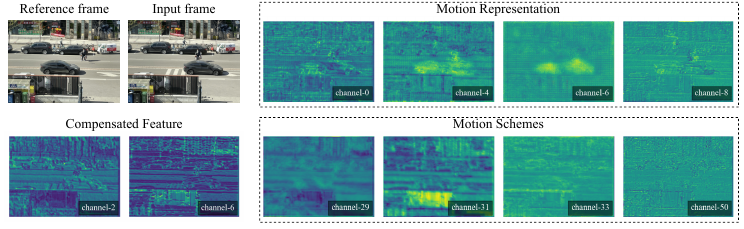}
\caption{\textcolor{black}{Visualization of the compensated features, motion representations, and motion schemes. }}
\label{fig_rebuttal_memc}   
\end{figure}

A visualization of our proposed scheme-based inter-prediction module is shown in Figure~\ref{fig_rebuttal_memc}. This module generates potential pattern schemes and then combines them through motion representation. The motion representation captures motion information, including local edge movements (e.g., channels 0 and 8) and large-scale motion (e.g., the rapid movement of the vehicle in channels 4 and 6). Moreover, motion schemes illustrate the potential components of the compensated feature, incorporating various types of pattern schemes. These schemes are subsequently synthesized into the compensated feature under the guidance of the motion representation. 

The compensated feature is a coarsely reconstructed feature obtained by inter-prediction and is similar to the current feature. The feature of the car is already moved to a new position in the compensated frame, as shown in Figure~\ref{fig_rebuttal_memc}. Since the compensated feature is merely a coarse-version feature of the current frame, the feature details are fulfilled through following perception-guided conditional coding process. 

\subsubsection{Relation Between Perception-Guided Conditional Coding and Spatial Redundancy Removal}

\begin{figure}[!h]
\centering
\includegraphics[width=0.80\linewidth]{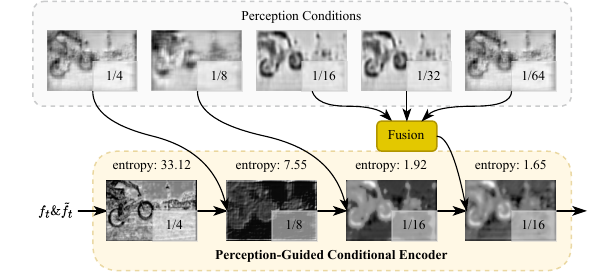}
\caption{\textcolor{black}{The process of perception-guided conditional encoding. The feature is compressed into a compact representation with lower entropy with the help of perception conditions as prior knowledge. }}
\label{fig_rebuttal_condition}   
\end{figure}

Spatial redundancy is prevalent in intermediate features, as adjacent regions often exhibit similar textures and high-frequency details, leading to overlapping or repetitive information. The perception-guided conditional coding module addresses this redundancy through two key perspectives:
First, as depicted in Figure~\ref{fig_rebuttal_condition}, the original features are downsampled multiple times and become smaller, more compact, and flatter. For a better understanding of the decrease in the amount of information, we take one frame as an example and calculate its entropy per pixel, as shown in equation~\ref{equation_entropy}. Figure~\ref{fig_rebuttal_condition} shows that the entropy of the feature decreases during the encoding process; then, the feature is compressed into a latent representation with a lower entropy that is suitable for entropy coding and transmission. 
Second, the intermediate features to be compressed have significant spatial structural correlations and repetition with the perception condition. Since the perception condition (acting as prior knowledge) is already accessible on both the encoder and decoder sides, there is no need to redundantly transmit this content from the encoder to the decoder. Instead, the decoder can effectively reconstruct the original content using the available perception information, further squeezing the spatial redundancy and enhancing the efficiency of the coding process. 

\begin{small}
\begin{equation}
    {entropy} = \sum^{N}_{i} p(f_i) log(p(f_i)) / (H \times W)
    \label{equation_entropy}
\end{equation}
\end{small}

\noindent where $N$ denotes the number of values in $f_i$, $p(\cdot)$ represents the probability of each value. $H$ and $W$ represent the height and width of the current frame, respectively. The feature $f$ undergoes 8-bit quantization for probability statistics. 

\textcolor{\revisioncolor}{\subsubsection{How the Multi-Branch Architecture of FST Works}}

\begin{figure}[!h]
\centering
\includegraphics[width=0.80\linewidth]{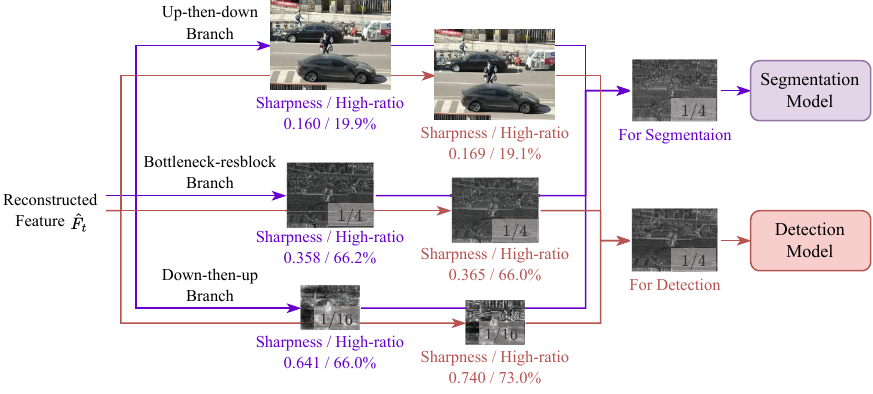}
\caption{\textcolor{\revisioncolor}{Visualization of the intermediate results produced by each branch in the FST module for video instance segmentation and object detection tasks. }}
\label{fig_revision_multibranch}   
\end{figure}

\begin{figure}[!h]
\centering
\includegraphics[width=0.80\linewidth]{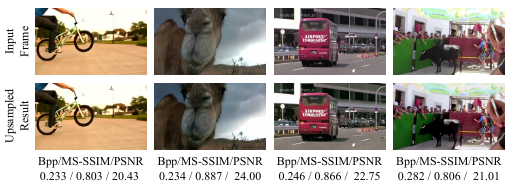}
\caption{Visualization of the upsampled results. The ``up-then-down" branch can coarsely reconstruct the original content in the pixel domain, which helps the feature transfer process gain knowledge of pixel-domain content.}
\label{fig_rebuttal_reconstruction}   
\end{figure}

\textcolor{\revisioncolor}{To better understand the behavior of the three branches contained in the FST module, intermediate results produced by each branch are visualized in Figure~\ref{fig_revision_multibranch}. Since each FST module is optimized for a specific task, different branches contribute differently depending on the given task. We use two metrics to evaluate the intermediate results: Sobel-based sharpness and the high-frequency ratio, defined as the proportion of FFT energy beyond the central $1/4$ area. First, in terms of the high-frequency ratio, the down-then-up branch tends to produce global feature maps with richer content for the coarse-grained detection task. In contrast, the up-then-down and bottleneck-resblock branches generate richer low-level details for the fine-grained instance segmentation task. This aligns with the task needs: segmentation relies more on local details to support pixel-level classification and fine boundary extraction. Second, from the perspective of sharpness, all three branches tend to produce smoother intermediate results for downstream instance segmentation tasks. This also matches the task requirements, as segmentation prefers spatial consistency and avoids discontinuities or jagged edges. 
Furthermore, the up-then-down branch enhances the feature-domain transformation by coarsely reconstructing the original frame, making the FST module aware of pixel-domain content, as illustrated in Figure~\ref{fig_rebuttal_reconstruction}. }

\textcolor{\revisioncolor}{\subsubsection{Robustness Analysis}}

\begin{figure}[!h]
\centering
\includegraphics[width=0.90\linewidth]{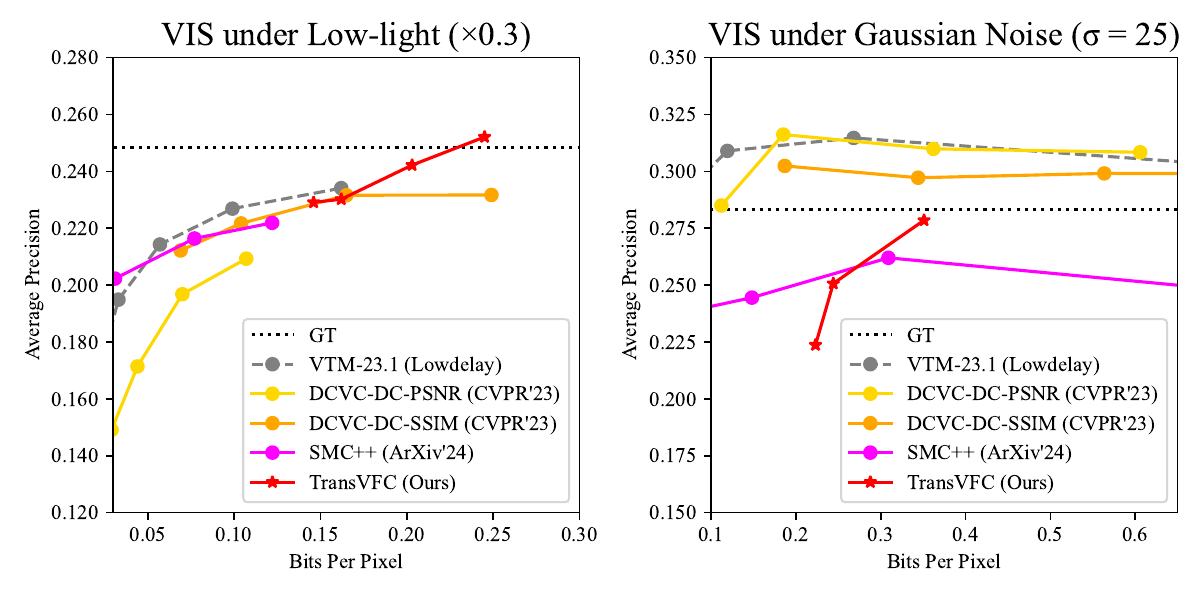}
\caption{\textcolor{\revisioncolor}{Rate-task performance of Video Instance Segmentation (VIS) in low-light and Gaussian noise scenarios. The proposed TransVFC performs well under low-light scenarios but is less effective when facing pixel-domain noise.}}
\label{fig_robustness}   
\end{figure}

\textcolor{\revisioncolor}{The proposed TransVFC framework is evaluated on the Video Instance Segmentation (VIS) task under two types of degradation: low-light and Gaussian noise. For the low-light scenario, following the existing approach~\cite{lolv1}, we map the videos to the YCbCr space, reduce the Y channel ($0.3\times$), and then convert them back to the RGB space. For the Gaussian noise scenario, we follow~\cite{zamir2022restormer} and add Gaussian noise with $\sigma=25$. As shown by the experimental results in Figure~\ref{fig_robustness}, our method maintains SOTA rate-task performance under low-light conditions. When Gaussian noise is introduced in the pixel domain, the bitrate of all methods increases significantly. Notably, HVS-oriented VTM and DCVC-DC can effectively suppress and filter such pixel-domain noise, and the reconstructed frames are mostly denoised during compression, resulting in task performance that even surpasses that of the uncompressed and noisy input. In contrast, our feature-compression-based method performs worse in this case, as pixel-domain noise harms the efficiency of feature compression. Addressing this limitation will be one of the directions of our future work. }

\subsection{Ablation Study}

Ablation experiments are conducted on the video instance segmentation task and the CrossVIS~\cite{crossvis} model. 

\subsubsection{Ablation on Video Feature Codec}

\begin{table}[!h]
\centering
\scriptsize \footnotesize
\begin{tabular}{ccccc}
\hline
\multirow{2}{*}{Models} & Scheme-based & Perception & Perception & \multirow{2}{*}{BD-Rate (\%)$\downarrow$}\\
 & inter-prediction & condition & loss & \\
\hline
Model 1 & \ding{55} & \ding{51} & \ding{51} & +11.37 \\
Model 2 & \ding{51} & \ding{55} & \ding{51} & +13.85 \\
Model 3 & \ding{51} & \ding{51} & \ding{55} & +36.92 \\
Model 4 & \ding{51} & \ding{55} & \ding{55} & +40.18 \\
Model 5 & \ding{55} & \ding{55} & \ding{55} & +46.71 \\
\hline
\end{tabular}
\caption{\textcolor{black}{Ablation study on the proposed components in the video feature codec. }}
\label{table_ablation_codec}
\end{table}

To verify the effectiveness of the proposed scheme-based inter-prediction, the proposed motion estimation and motion compensation modules are replaced with the existing deformable-convolution-based approach~\cite{fvc}; this model, which is represented as Model 1 in Table~\ref{table_ablation_codec}, results in an 11.37\% average bitrate increase. To verify the effectiveness of the perception conditions, we retain the framework structure but do not use $C_{enc}$ and $C_{dec}$ as conditions, named Model 2. It is demonstrated that reconstructing video features without the conditions causes a 13.85\% bitrate increase. Furthermore, the high-level perception loss $D_{p}$ is removed in Model 3, resulting in a 36.92\% bitrate increase. The result of Model 4 indicates that introducing high-level perception in both the conditional coding process and the loss function can significantly increase the rate-task performance (40.18\% in total). Additionally, when both scheme-based inter-prediction and perception-guided conditional coding modules are removed (Model 5), simplifying the codec to a structure similar to FVC~\cite{fvc} with deformable-convolution-based inter-prediction and residual coding, the bitrate increases by 46.71\%. 

\subsubsection{Ablation on Feature Space Transform Module}

\begin{table}[!h]
\centering
\scriptsize \footnotesize
\begin{tabular}{ccccc}
\hline
Models & Bottleneck-resblock & Down-then-up & Up-then-down & BD-Rate(\%)$\downarrow$ \\
\hline
Model 6 & \ding{51} & \ding{51} & \ding{55} & +2.83 \\
Model 7 & \ding{51} & \ding{55} & \ding{51} & +2.69 \\
Model 8 & \ding{51} & \ding{55} & \ding{55} & +5.69 \\
\hline
\end{tabular}
\caption{Ablation study on the proposed FST module. }
\label{table_ablation_fst}
\end{table}

To verify the function of each branch contained in the FST module, we remove the up-then-down branch (Model 6), the down-then-up branch (Model 7), and both branches (Model 8), as shown in Table~\ref{table_ablation_fst}. The experimental results demonstrate that each branch plays a significant role in the quality of the feature space transformation. 

Additionally, we conduct an ablation study on the complexity of the FST module. We roughly double the number of parameters of the FST. The number of parameters in FST increases from 4.30M to 5.92M (+37\%), and the MACs per pixel rise from 0.14M to 0.26M (+86\%). The FST further reduces the BD-Rate by only 2.42\%. We also roughly reduce the number of res-blocks. The number of parameters is reduced to 3.53M and MACs per pixel reduce to 0.12M. The BD-Rate increases by 5.60\%. This finding shows that the current structure is appropriate since greater complexity results in only a limited BD-Rate reduction.

\subsubsection{Comparison Among Different Approaches in ATC Paradigm}

\begin{table}[!h]
\setlength\tabcolsep{2pt} 
\centering
\scriptsize \footnotesize
\begin{tabular}{ccccccc}
\hline
\multirow{2}{*}{Models}     & \multirow{2}{*}{Codec}    & \multirow{2}{*}{Task} & \multirow{2}{*}{BD-Rate(\%)$\downarrow$}  & Optimzed  & GPU mem       & Training time \\
                        & &  &                           & params    & (GiB) & per step (s)\\
\hline
Model 9 & \ding{51} & \ding{55} & -6.33 & 22.4M & 19.3(+12.9\%) & 1.430(+14.1\%) \\
Model 10 & \ding{55} & \ding{51} & -7.16 & 37.4M & 18.6(+8.8\%) & 1.374(+9.7\%) \\
Ours & \ding{55} & \ding{55} & 0 & 4.3M  & 17.1 & 1.253 \\
\hline
\end{tabular}
\caption{Ablation study on different approaches in ATC paradigm. ``\ding{51}'' means optimized and ``\ding{55}'' means frozen.}
\label{table_atc}
\end{table}

Other ATC-based VCM pipelines are implemented based on TransVFC, as detailed in Table~\ref{table_atc}. Referring to \cite{misra2022video, shao2020bottlenet++, sheng2023lvvc}, we fine-tune either the upstream video feature codec (Model 9) or the downstream task network (Model 10) instead of the FST module. The experimental results indicate that training either the upstream or the downstream network leads to additional bitrate savings. However, this comes at the cost of needing to optimize more parameters, consuming more computational resources and training time. Benefiting from the FST module, our approach uses fewer computational resources and avoids redeploying the upstream video feature codec or downstream task networks, offering better scalability.

\subsubsection{Influence on I-Frame Codec}

The proposed TransVFC directly uses the feature of the first lossless frame and calculates the bpp value of its original I-frame jpeg file. Additionally, our experiments show that introducing x265 for I-frame compression causes a 5.10\% bitrate increase.

\section{Conclusion}

We propose a TransVFC framework. It offers a scalable solution for multitask VCM scenarios and eliminates the need for fine-tuning the upstream codec and the downstream machine vision tasks. We devise a novel neural-based video feature codec to achieve continuous feature compression; this method incorporates a scheme-based inter-prediction module for feature-domain temporal redundancy squeezing and employs perception-guided conditional coding to make the features better align with machine perception. We designe an FST module to effectively transfer the intermediate features to multiple downstream tasks. Experiments are conducted on three downstream machine vision tasks at different granularities, demonstrating that TransVFC delivers promising compression efficiency and scalability. 

Despite these promising results, our approach has limitations. Its performance tends to decrease in low-bitrate scenarios. This decrease may stem from inter-prediction challenges when addressing low-quality features, which introduces cumulative error in the feature domain and affects the overall rate-task performance. \textcolor{\revisioncolor}{In addition, our method exhibits limited robustness when confronted with Gaussian noise in the pixel domain.} Moreover, a gap remains between our framework and real-time systems such as FFmpeg~\cite{x265}. \textcolor{\revisioncolor}{In future work, we plan to increase the performance of our method under low-bitrate constraints, reduce its coding latency, explore new prior utilization and feature fusion strategies, improve robustness on degraded input video, and incorporate a variable-bitrate mechanism.}

We hope that our approach can inspire advancements in video feature compression for multitask scenarios and contribute to the development of ATC-based VCM methods.

\section{Acknowledgments}

This work is supported by the National Natural Science Foundation of China (62120106009, 62372036, and U24B20179).

\section{Declaration of Generative AI and AI-assisted Technologies in the Writing Process.}

During the preparation of this work, the author(s) used ChatGPT to polish the manuscript and enhance its readability. After using this tool/service, the author(s) reviewed and edited the content as needed and take(s) full responsibility for the content of the published article.


\end{document}